
%

%
\font\eightrm=cmr8
\font\eighti=cmmi8
\font\eightsy=cmsy8
\font\eightbf=cmbx8
\font\eighttt=cmtt8
\font\eightit=cmti8
\font\eightsl=cmsl8
\font\sixrm=cmr6
\font\sixi=cmmi6
\font\sixsy=cmsy6
\font\sixbf=cmbx6
\catcode`@11
\newskip\ttglue
\font\grrm=cmbx10 scaled 1200

\def\eightpoint{\def\rm{\fam0\eightrm}
\textfont0=\eightrm \scriptfont0=\sixrm \scriptscriptfont0=\fiverm
\textfont1=\eighti \scriptfont1=\sixi \scriptscriptfont1=\fivei
\textfont2=\eightsy \scriptfont2=\sixsy \scriptscriptfont2=\fivesy
\textfont3=\tenex \scriptfont3=\tenex \scriptscriptfont3=\tenex
\textfont\itfam=\eightit \def\it{\fam\itfam\eightit}
\textfont\slfam=\eightsl \def\sl{\fam\slfam\eightsl}
\textfont\ttfam=\eighttt \def\tt{\fam\ttfam\eighttt}
\textfont\bffam=\eightbf
\scriptfont\bffam=\sixbf
\scriptscriptfont\bffam=\fivebf \def\bf{\fam\bffam\eightbf}
\tt \ttglue=.5em plus.25em minus.15em
\normalbaselineskip=6pt
\setbox\strutbox=\hbox{\vrule height7pt width0pt depth2pt}
\let\sc=\sixrm \let\big=\eightbig \normalbaselines\rm}
\newinsert\footins
\def\newfoot#1{\let\@sf\empty
  \ifhmode\edef\@sf{\spacefactor\the\spacefactor}\fi
  #1\@sf\vfootnote{#1}}
\def\vfootnote#1{\insert\footins\bgroup\eightpoint
  \interlinepenalty\interfootnotelinepenalty
  \splittopskip\ht\strutbox 
  \splitmaxdepth\dp\strutbox \floatingpenalty\@MM
  \leftskip\z@skip \rightskip\z@skip
  \textindent{#1}\footstrut\futurelet\next\fo@t}
\def\fo@t{\ifcat\bgroup\noexpand\next \let\next\f@@t
  \else\let\next\f@t\fi \next}
\def\f@@t{\bgroup\aftergroup\@foot\let\next}
\def\f@t#1{#1\@foot}
\def\@foot{\strut\egroup}
\def\footstrut{\vbox to\splittopskip{}}
\skip\footins=\bigskipamount 
\count\footins=1000 
\dimen\footins=8in 

\def\ref#1{$^{#1}$}
\def\flex{\raise 6pt\hbox{$\leftrightarrow $}\! \! \! \! \! \! }
\def\oversome#1{ \raise 8pt\hbox{$\scriptscriptstyle #1$}\! \! \! \! \! \! }
\def\tr{ \mathop{\rm tr}}

\newbox\bigstrutbox
\setbox\bigstrutbox=\hbox{\vrule height10pt depth5pt width0pt}
\def\bigstrut{\relax\ifmmode\copy\bigstrutbox\else\unhcopy\bigstrutbox\fi}
\def\refer[#1/#2]{ \item{#1} {{#2}} }
\def\rev<#1/#2/#3/#4>{{\it #1\/} {\bf#2}, {#3}({#4})}
\def\boxit#1{\vbox{\hrule\hbox{\vrule\kern3pt
\vbox{\kern3pt#1\kern3pt}\kern3pt\vrule}\hrule}}

\def\2figure#1#2#3#4{\vbox{ \hrule width#1truecm \hbox{\vrule height#2truecm
\hskip #1truecm
\vrule height#2truecm }\hrule width#1truecm \hbox{\vrule\vbox{\hsize #1truecm
\baselineskip=10pt
\noindent\strut#3}\vrule}\hrule width#1truecm
\hbox{\vrule\vbox{\hsize #1truecm
\baselineskip=10pt
\noindent\strut#4}\vrule}\hrule width#1truecm  }}
\def\3figure#1#2#3#4#5{\vbox{ \hrule width#1truecm \hbox{\vrule height#2truecm
\hskip #1truecm
\vrule height#2truecm }\hrule width#1truecm \hbox{\vrule\vbox{\hsize #1truecm
\baselineskip=10pt
\noindent\strut#3}\vrule}\hrule width#1truecm
 \hbox{\vrule\vbox{\hsize #1truecm
\baselineskip=10pt
\noindent\strut#4}\vrule}
\hrule width#1truecm \hbox{\vrule\vbox{\hsize #1truecm
\baselineskip=10pt
\noindent\strut#5}\vrule}\hrule width#1truecm  }}

\def\sqr#1#2{{\vcenter{\hrule height.#2pt
   \hbox{\vrule width.#2pt height#1pt \kern#1pt
    \vrule width.#2pt}
    \hrule height.#2pt}}}
\def\dal{\mathchoice{\sqr{6}{4}}{\sqr{5}{3}}{\sqr{5}3}{\sqr{4}3} \, }


\def\smin{\,\raise 0.06em \hbox{${\scriptstyle \in}$}\,}
\def\smsubset{\,\raise 0.06em \hbox{${\scriptstyle \subset}$}\,}

\def\Natural{\hbox{\hskip 1.5pt\hbox to 0pt{\hskip -2pt I\hss}N}}

\def\Rational{\hbox{\hbox to 0pt{\hskip 2.7pt \vrule height 6.5pt
                                  depth -0.2pt width 0.8pt \hss}Q}}
\def\Real{\hbox{\hskip 1.5pt\hbox to 0pt{\hskip -2pt I\hss}R}}
\def\Complex{\hbox{\hbox to 0pt{\hskip 2.7pt \vrule height 6.5pt
                                  depth -0.2pt width 0.8pt \hss}C}}

\def \E {{{\rm e}}}


\def \1ok{{1\over \kappa ^2} }

\def \3dslim {{\rm DS}\!\!\!\!\!\!\!\!\lim }
\def \4dslim {{\rm DS}\!\!\!\!\!\!\!\!\!\!\lim }
\def \tr {{\rm tr}\, }
\def \ln {{\rm ln}\, }
\def \2kk{\left( \matrix {2k\cr k\cr }\right) }
\def \Rs4{{R^k\over 4^k} }

\def \1ok{{1\over \kappa ^2} }



\magnification 1200
\def \E {{\rm e}}
\nopagenumbers

\vskip .3cm
\centerline{\grrm Three-dimensional quantum electrodynamics }
\vskip .3cm
\centerline {\grrm as an effective interaction\newfoot {${}^*$}{This
work is supported in part by funds provided by the U.S. Department of
Energy(D.O.E.) under cooperative agreement \#DF-FC02-94ER40818.} }
\vskip 1cm
\centerline {E. Abdalla\newfoot {${}^{\dag}$}{Email: elcio@ictp.trieste.it,
abdalla@surya11.cern.ch. Permanent address: Instituto de
F\'\i sica - USP, C.P. 20516, S. Paulo, Brazil.}}
\vskip .5truecm
\centerline {CERN, Theory Division, CH-1211 Geneva 23, Switzerland}
\vskip .2truecm
\centerline {International Centre for Theoretical Physics - ICTP}
\centerline {34100 Trieste, Italy}
\vskip .3truecm
\centerline {and}
\vskip .3cm
\centerline{F. M. de Carvalho Filho\newfoot{${}^{\ddag}$}{Email:
farnezio@mitlns.mit.edu.
Permanent Address: Ins\-ti\-tuto de Ci\^encias, Escola Federal
de Engenharia de Itajub\'a, C.P. 50, Itajub\'a, CEP: 375000-000, Minas Gerais,
Brazil.}}
\vskip .5truecm
\centerline {Center for Theoretical Physics}
\centerline{Laboratory for Nuclear Science and Department of Physics}
\centerline{Massachusetts Institute of Technology}
\centerline{Cambridge, Massachusetts 02139 U.S.A.}

\vskip 2cm
\centerline{\bf Abstract}
\vskip .5cm
\noindent
We obtain a Quantum Electrodynamics in 2+1 dimensions by applying a
Kaluza--Klein type method of dimensional reduction to Quantum
Electrodynamics in 3+1 dimensions rendering the model more realistic to
application in solid-state systems, invariant under translations in one
direction. We show that the model obtained leads to an effective action
exhibiting an interesting phase structure and that the generated
Chern--Simons term survives only in the broken phase.
\vskip .5cm
\noindent PACS number(s): 11.10.Kk, 12.20.-m, 73.40.Hm

\vfill
\noindent MIT--CTP--2488

\noindent IC/95/350

\noindent hep-th/9511132
\hfill November 1995

\vskip 1cm
\eject
\countdef\pageno=0 \pageno=1
\newtoks\footline \footline={\hss\tenrm\folio\hss}
\def\folio{\ifnum\pageno<0 \romannumeral-\pageno \else\number\pageno \fi}
\def\advancepageno{\ifnum\pageno<0 \global\advance\pageno by -1
\else\global\advance\pageno by 1 \fi}
\centerline {\bf 1. Introduction}
\vskip .7cm

\noindent Models in (2+1) dimensions with or without Chern--Simons terms have
been a subject of an intense theoretical study in the last years\ref{1-9}, for
several important reasons. First, they are generalizations of
two-dimensional models\ref{10}, where one can have an interesting theoretical
laboratory for check of ideas, a little more complex than the two-dimensional
case, but still without all the technical complexity of four dimensions. This
is the case of the gravitation theory in (2+1) dimensions\ref{1} that is very
simple from the classical point of view: it has zero degrees of freedom, but
topological effects appear, and are effective to the understanding of the
physics involved. Another equally important to study
such models, is that they can be useful in determining properties of certain
important practical condensed-matter systems. In particular
many efforts have been recently devoted to find good
theoretical explanations for the high-T$_c$ superconductivity\ref{6-9,11,12}
as well as
for the fractional quantum Hall effect\ref{13} which are observed in some
layered ceramics. These materials exhibit a planar geometry that can,
in principle, be described by models in quantum field theory with only two
spatial coordinates. So one of the main challenges in this area
is to find correct models for
description of these phenomena.

In the present work we obtain a special model,
the quantum electrodynamics in (2+1) dimensions with a Chern--Simons term,
starting from the quantum electrodynamics in (3+1)-dimensional space time
(QED$_4$), which we believed to be a good choice as starting point since it is
always the most realistic theory when the effects previously referred to are
treated as in the usual way. We obtain the model via a Kaluza--Klein-type
mechanism for periodic fields\ref{14-17}, in which a dimensional reduction of
extra coordinate is made. Such an extra dimension is associated to the
direction of the layers in the sample in the same way as Bloch wave functions
are found in solid-state systems, and we are naturally led to a compact
space (one torus in this case) such that the
three-dimensional physics is a low energy approximation of a broader theory.
Indeed such reductions are commonly used in condesend-matter systems such
as the fractional Hall effect, where the edges currents will be described
by an even simpler model, in two-dimensional space-time.(See,for instance,
Ref.[18] and the list of references therein.)
We also analyse in this work the phase structure of the three-dimensional
quantum electrodynamics obtained(KKB-QED$_3$) with special attention
being paid to the role played by the Chern--Simons term.

The paper is organized as follows: In section 2 we start with the
four-dimensional quantum electrodynamics and obtain a theory in three
dimensions via the above mentioned Kaluza--Klein scheme for periodic
fields(KKB). In section 3 we obtain the
effective
potential for this three-dimensional quantum electrodynamics and find its
minimum. In section 4 we investigate the phase structure of the model and
calculate the Chern--Simons term induced by the fermions. Section 5
contains our conclu\-sion\-s.

\vskip 2.2cm
\penalty-300
\centerline {\bf 2. The KKB technique and the model}
\vskip .7cm
\nobreak
\noindent In order to obtain a model in three dimensions, we start with the
quantum electrodynamics in (3+1) dimensions described by the Lagrangian
$$
{\cal L}= \overline \Psi i D \Psi - M\overline \Psi \Psi - {1\over
4}F_{mn}F^{mn}\quad ,\eqno(2.1)
$$
where
$$
\eqalign{
i D& = \Gamma^m (i \partial_m + e A_m)= i \Gamma^m D_m\quad , \quad m,n = 0,
1, 2, 3\quad ,\cr
i D_\mu & = i \partial _\mu + e A_\mu \quad ,\quad \mu = 0, 2, 3 \quad .\cr}
$$

\noindent $M$ and $e$ are the mass and the electric charge in four dimensions
and $\Gamma^m$ are the Dirac gamma matrices. The gauge field strength $F^{mn}$
is given in terms of the vector potential $A_m$ by $F^{mn}= \partial
^mA^n-\partial ^nA^m$. We employ the chiral representation of Dirac $\Gamma$
matrices, i.e.,
$$
\Gamma^0 = \pmatrix {0&-1\cr -1&0\cr}\quad ,\quad \vec \Gamma=  \pmatrix
{0&\vec\sigma\cr -\vec\sigma&0\cr}\quad .
$$
They satisfy the Clifford algebra $\{ \Gamma^m,\Gamma^n\} = 2G^{mn}$, with
the metric tensor $G^{mn} = (+ - - -)$.

The Dirac spinor is given in terms of the Pauli spinors, by
$$
\Psi = \pmatrix {\psi\cr \chi\cr} \quad ,\quad \overline \Psi = \psi^\dagger
\Gamma^0 = \pmatrix {-\chi^\dagger & -\psi^\dagger\cr}\quad .
$$

In terms of the three-dimensional quantities, now represented by greek indices
$(\mu, \nu = 0, 2, 3)$ and small letters, the Lagrangian is rewritten, as
$$
\eqalignno{
{\cal L} = \, & i \overline \psi \gamma^\mu D_\mu \psi + i \overline \chi
\gamma^\mu D_\mu \chi - {1\over 4}F_{\mu\nu}F^{\mu\nu} + {1\over 2}
\partial^\mu A_1\partial_\mu A_1 + \cr
& + {1\over 2} (\partial _1A_0)^2 - {1\over 2} (\partial _1A_2)^2 -
{1\over 2}(\partial_1A_3)^2
-\partial_1A_0\partial_0A_1+\partial_1A_3\partial_3A_1 + \cr
& +\partial _1A_2\partial_2A_1 + M(\overline \chi \psi + \overline \psi
\chi) + e
(\overline \psi \psi A_1 - \overline \chi \chi A_1) + (\overline \psi
\partial _1\psi - \overline \chi \partial _1 \chi)\quad ,&(2.2)\cr}
$$
where the index 1 is compactified.

Above, the representation of Dirac $\gamma$ matrices in three dimensions has
Pauli matrices given by
$$
\gamma_0 =\left( \matrix{0&1\cr 1&0\cr}\right) = \gamma^{"5"} = \sigma^1\quad
;\quad \gamma_2 =\left(\matrix{0&-1\cr 1& 0\cr}\right) = - i \sigma^3\quad
;\quad \gamma_3 =\left(\matrix{i&0\cr 0&-i\cr}\right) = i \sigma ^2\quad ,
$$
which, in particular, satisfy the useful property $\gamma^\mu\gamma^\nu =
g^{\mu\nu} - i \epsilon^{\mu\nu\rho}\gamma_\rho$. Where $\epsilon_{\mu\nu\rho}$
is the Levi-Civita tensor which, in three dimensions, is totally antisymmetric.

Now we implement the KKB scheme\ref{14} by considering that the model
admits a spontaneous compactification\ref{14-17} on a Minkowski
(2+1)-dimensional space-time and $G/H$ a homogeneous compact space associated
to the extra coordinate (one torus in this case) that is assumed to be a
circle of length $l_x=l$. Physically, this is equivalent to suppose that we
are dealing with a sample with layers, whose interspace $l$ is very small,
and physics is periodic in $x\to x +l$. In  condensed-matter
systems, $l$ is the cristal interlayer space. We split the four components of
the four-vector position in $x_m= (x_\mu, x_1) = (x_\mu, x)$ and make the
decomposition of the vector gauge field in $A_m \to (A_\mu, A)$, where we
denote $A_x=A_1=A$.

Furthermore, we also assume:
$$
\eqalign{
\psi(x_\mu, x+l) & = \psi (x_\mu, l)\quad ,\cr
\chi(x_\mu, x+l) & = \chi (x_\mu, x)\quad ,\cr
A(x_\mu, x+l) & = A (x_\mu, x)\quad .\cr}
$$

The above relations  break the (3+1)-\-di\-mens\-ional Poincar\'e invariance
of (2.1), $P_{3+1}$ to a (2+1)-\-di\-mens\-ional Poincar\'e invariance times
$U(1)$, that is, $P_{3+1} \to P_{2+1} \times U(1)$.

Now $\psi, \, \chi$ and $A$ can be expanded in Fourier series (we are
admitting planar structures in specified matter-condensed systems of interest
in which the Bloch theorem is valid).
$$
\eqalignno{
\psi(x_\mu, x) = & {1\over \sqrt l} \sum_r \psi^{(r)} \E^{{irx\over l}}\quad ,
\cr
\chi(x_\mu, x) = & {1\over \sqrt l} \sum_r \chi^{(r)} \E^{{irx\over l}} \quad ,
\cr
A_\mu (x_\mu, x) = & {1\over\sqrt l}\sum_r A_\mu^{(r)}\E^{{irx\over l}}\quad ,
&(2.3)\cr}
$$
with $r$ integer.

By substituting the above expansions in the Lagrangian (2.2) we can integrate
over $x$ from zero to $l$ and obtain a three-dimensional description of this
theory through the reduced Lagrangian.

$$
\eqalignno{
\widetilde {\cal L} & =  i \overline \psi \gamma^\mu D_\mu \psi + i
\overline  \chi \gamma^\mu \tilde D_\mu \chi  + M(\overline \chi \psi +
\overline \psi\chi) +
qA(\overline \psi \psi - \overline \chi\chi) -
{1\over 4}F_{\mu\nu}F^{\mu\nu}+{1\over 2}\partial^\mu A\partial_\mu A - \cr
& -\sum_rm_{(r)}\left(\overline\psi^{(-r)}\psi^{(r)}-\overline\chi^{(-r)}
\chi^{(r)}\right)-i \sum _r m_{(r)} \left(A^{(-r)}\partial^\mu A_\mu^{(r)}
\right) + {1\over 2} \sum _r m^2_{(r)}A_\mu^{(-r)}A^{\mu(r)}\cr
{} & {}&(2.4)\cr}
$$
where $ m^2_{(r)}= {r^2\over l^2}$, and the momentum of $\chi$ is opposite to
that of $\psi$ (see notation after eq. (3.5)).

The Lagrangian $\widetilde {\cal L}$ is adentical to ${\cal L}$ up to some
rescaling of the coupling constants and the fields in order to give them the
canonical dimensions of the QED$_4$. These rescaling factors disappear after
integration over $x$. Notice that there are two limiting cases: if $l \to
\infty$, the original QED$_4$ is recovered; if $l\to 0$ ($l \sim $ size
of the
compact space or torus), we are making, in fact, an ordinary dimensional
reduction in which the fields do not depend on the extra-coordinate $x$. In
condensed-matter systems with layered structures, such as high-T$_c$
superconductors, the truncation by letting $l\to 0$ is very reasonable since
$l$ will have to be related to the interlayers distance ($\sim 10^{-9}$m)and
to the effective
mass $m\sim {1\over l} $. In these cases, $l << l_y $ and $l << l_z$.

\vskip 1cm
\penalty-300
\centerline {\bf 3. The effective potential in the KKB-QED$_3$}
\vskip .7cm
\nobreak
\noindent In order to study the physics of the KKB-QED$_3$ with fermions
inducing a parity-violating Chern--Simons term we now calculate the
effective action upon integration over the fer\-mions. The term
$$
 \sum_r m_{(r)}\left(\overline \psi^{(-r)} \psi^{(r)} - \overline \chi^
{(-r)}\chi^{(r)}\right) \quad,
$$
can be neglegted since in three dimensions an even number of fermions can be
paired to form Dirac fermions with parity-conserving mass terms\ref{4}. The
two last terms can be dropped because when we make calculations in high
orders in perturbation theory all higher-order graphs in the effective
action are proportional to powers of ${1\over m}$ or ${1\over r/l}$ and
they vanish when $m\to \infty$, or $l\to 0$.

Thus, our reduced Lagrangian becomes, in this approximation
$$
\eqalignno{
{\cal L} = &i\overline\psi\gamma^\mu\partial_\mu\psi+i\overline\chi\gamma^\mu
\tilde\partial_\mu\chi+M(\overline\chi\psi+\overline\psi\chi)+qA(\overline\psi
\psi - \overline \chi\chi) - {1\over 4} F_{\mu\nu}F^{\mu\nu} \cr
& +{1\over 2}\partial^\mu A\partial_\mu A+q\left(\overline\psi\gamma_\mu\psi
A^\mu + \overline \chi \gamma_\mu \chi A^\mu \right) \quad ,&(3.1)\cr}
$$
where $q$ is an effective electric charge in three dimensions. Now, we may work
with the above Lagrangian to obtain, in perturbation theory, a renormalized
effective gauge field action which contains a Chern--Simons induced by
fermions.
This term will have to appear in the effective action when ultraviolet
divergences are regulated in a gauge-invariant way, for example, using the
Pauli-Villars regularization. Only the vacuum polarization and triangle graphs
are ultraviolet divergent. But in the Abelian case, only the first one requires
regularization.

Thus, we integrate over the fermions fields and calculate the effective
potential $V_{eff}(A)$ at ono-loop order. Before, making the substitution $A\to
A' = A +a$, where $a= \langle 0\vert A\vert 0\rangle = \langle A \rangle $ we
have
$$
\eqalignno{
{\cal L}_{eff} = & i \overline \psi \gamma^\mu \partial_\mu \psi + i \overline
\chi \gamma^\mu\tilde\partial_\mu \chi  + M(\overline \chi \psi + \overline\psi
\chi) + qA(\overline \psi \psi - \overline \chi\chi)
+ qa(\overline \psi \psi - \overline \chi\chi)\cr
&-{1\over 4} F_{\mu\nu}F^{\mu\nu} + {1\over 2} \partial^\mu A\partial _\mu
A + q\left( \overline \psi\gamma_\mu
\psi+ \overline \chi \gamma_\mu \chi \right)A^\mu \quad ,&(3.2)\cr}
$$

The effective potential $V_{eff} (a) = i \ln \det {\cal O}$, or, in  a more
conveniente way,
$$
V_{eff} = - i \E^{\tr \ln {\cal O}} \quad ,\eqno(3.3)
$$
where
$$
{\cal O} = \pmatrix {i\partial - qa & M \cr M & i\partial + qa\cr}=
\pmatrix{k_0-\vec k\cdot \vec \sigma -\mu \sigma_1& M\sigma_1\cr
M\sigma_1 & k_0+\vec \sigma\cdot\vec k+\mu \sigma_1\cr} \quad . \eqno(3.4)
$$

After some algebra, $V_{eff}$ is given by
$$
\eqalignno{
V_{eff}(a) = & - {i\over 2} \tr \Big[ \int {{\rm d}^3 k \over (2\pi)^3} \ln
\left( 1+ {(qa)^2\over (k^2-M^2)}\right) 1 - \cr
& - \int {{\rm d}^3 k \over (2\pi)^3} \sum _n {(qa)^{2n+1}\over (2n+1)}
{1\over
(k^2-m^2)} \pmatrix {\not \!k&M\cr M&\tilde {\not \!k}\cr} \Big]\quad,
&(3.5)\cr}
$$
where $\not \! k = k_0 - \vec \sigma \cdot \vec k$ and $\tilde {\not \! k}=
k_0 + \vec\sigma\cdot \vec k$.

Making a Wick rotation, and taking the trace, we have
$$
V_{eff} =  \int {{\rm d}^3 k \over (2\pi)^3} \ln \left( 1 - {(qa)^2\over
(k^2 + M^2)}\right) \quad ,\eqno(3.6)
$$
and it can be rewritten in terms of a cutoff $\Lambda$
$$
\eqalignno{
V_{eff}(a)=& {1\over 2\pi^2} \int _0^\Lambda {\rm d} k k^2 \left[ \ln (k^2 +
M^2 - (qa)^2) - \ln (k^2 + M^2) \right]\cr
=& - {1\over 2\pi^2} \left[ \Lambda (qa)^2 - {\pi\over 3} (qa)^3 \right] +
{\cal O}\left( {1\over \Lambda}\right)\quad.&(3.7)\cr}
$$
Notice that, above, the cutoff $\Lambda$ can be removed through a
renormalization.

By analysing the above  effective potential, we obtain two values that make it
minimum $\langle qa\rangle = qa = \mu =0$ and $\mu = {2\Lambda\over \pi}$.
Therefore, our system, described by a reduced quantum electrodynamics,
the KKB-QED$_3$,
exhibits a
phase transition, and there is a phase in which the field acquires
mass. Furthermore, if we integrate out completely the $A$ field, we find also a
Gross-Neveu interaction in three dimensions.

\vskip 1cm
\penalty-300
\centerline{\bf 4. The Chern--Simons term and the phase structure}
\vskip .7cm
\nobreak
\noindent Now that we have shown the existence of two phases in our theory, the
next question to address is whether a Chern--Simons term is a given phase and
which is its role.

First we find the fermion propagators of the KKB-QED$_3$. Formally, in terms
of the
path integral with respect to the fields $\psi, \overline \psi, \chi, \overline
\chi, A_\mu$ and $A$, the generating functional is
$$
{\cal Z} [J, \overline J, \eta, \overline \eta, K, j] = \int {\cal D}\overline
\psi {\cal D}\psi  {\cal D}\overline \chi {\cal D} \chi {\cal D} A_\mu {\cal D}
A \, \E^{i\int {\rm d}^3 x \left[ {\cal L}_{eff} + (\overline J\psi + \overline
\psi
J + \overline \chi \eta + \overline \eta \chi) \right]}\quad ,\eqno(4.1)
$$
where ${\cal L}_{eff}$ is that one given by eq. (3.2) and $\overline J, J,
\eta, \overline \eta, K$ and $j$ are external c-numbers sources coupled to the
fields. The propagators read
$$
\eqalignno{
\langle \psi \overline \psi \rangle = & - {\delta ^2 {\cal Z} \over \delta
J\delta\overline J}\Big\vert_{J=\overline J=0} = S_{11}(k) = i
{\not \! k-\mu\over k^2- (M^2 + \mu^2)}\quad ,\cr
\langle \chi \overline \chi \rangle = & - {\delta ^2 {\cal Z} \over \delta
\eta\delta\overline \eta}\Big\vert_{\eta=\overline \eta=0} = S_{22}(k) =
i {\tilde {\not \! k}+\mu\over k^2- (M^2 + \mu^2)}\quad ,\cr
\langle \psi \overline \chi \rangle = & - {\delta ^2 {\cal Z} \over \delta
\overline J\delta\eta}\Big\vert_{\overline J=\eta = 0} = S_{12}(k) =
i {M\over k^2-(M^2 + \mu^2)}\quad ,\cr
\langle \chi \overline \psi \rangle = & - {\delta ^2 {\cal Z} \over \delta
\overline \eta\delta J}\Big\vert_{\overline \eta =J=0} = S_{21}(k) =
i {M\over k^2- (M^2 + \mu^2)}\quad .&(4.2)\cr}
$$

Defining $m^2_t=\mu^2 + M^2$, we compute below all contributions to the
effective action from vacuum polarization graphs, taking into account the above
propagators.

We can write the vacuum polarization as
$$
\Pi^{\mu\nu} = \Pi^{\mu\nu}_{11} + \Pi^{\mu\nu}_{22} + \Pi^{\mu\nu}_{12} +
\Pi^{\mu\nu}_{21}\quad .\eqno(4.3)
$$

The graphs involving only one kind of fermion ($\psi$ or $\chi$) give, for
example
$$
\Pi^{\mu\nu}_{11} = i q^2 \int {{\rm d}^3 k\over (2\pi)^3} {\tr
\left[\gamma^\mu (\not \! k-\mu)\gamma^\nu (\not \! k+\not
\! p-\mu)\right]\over (k^2- m_t^2)[(k+p)^2-m_t^2]}\quad .\eqno(4.4)
$$

Upon evaluating the trace over Dirac matrices and using the same regularization
prescription of ref. [3]  and [4], we obtain a contribution for the
Chern--Simons (CS) term only from
 $$
{\Pi^{\mu\nu}_{11}}_{CS}={\Pi^{\mu\nu}_{22}}_{CS}={-\mu\over\sqrt{M^2+\mu^2}}
{q^2\over 4\pi} (i\epsilon^{\mu\nu\rho} p_\rho)\quad .\eqno(4.5)
$$

This way, we have a Chern--Simons term generated that can be expressed in the
effective Lagrangian of the KK-QED$_3$ model. The effective Lagrangian is
given by the expression
$$
\eqalignno{
\widetilde {\cal L}_{eff} = & i \overline \psi \gamma^\mu \partial_\mu \psi +
i \overline \chi \gamma^\mu \tilde \partial_\mu \chi  + M(\overline \chi
\psi +
\overline\psi \chi) + \mu(\overline \psi \psi - \overline \chi\chi) -
{1\over 4} F_{\mu\nu}F^{\mu\nu}\cr
& + {1\over 2} \partial^\mu A\partial _\mu
A + q\left( \overline \psi\gamma_\mu A^\mu\psi+ \overline \chi
\gamma_\mu A^\mu \chi\right) + {q^2\over 4\pi} {\mu\over \vert m_t
\vert} \epsilon^{\mu\nu\rho} A_\mu \partial _\nu A_\rho + {\cal O} \left(
{1\over m_t}\right)\quad ,\cr
{}& {} &(4.6)\cr}
$$

By analysing this special Chern--Simons term, we notice that it vanishes when
the system is in the phase where the $A$ field is massless (either with
massive or massless fermions). However, at the broken phase, this term becomes
very important in the dynamics of the KK-QED$_3$ because it does not vanish
even
when the fermions are massive, except of course, if $M\to \infty$. When $M=0$,
we have $m_t=\mu$ and there will be a $\pm$ sign from $\mu/\vert m_t\vert$.

We can use eq. (4.6) to derive the vector current in terms of the
Chern--Simons term. We have
$$
q \langle j^\mu (x)\rangle = q\langle \overline \psi_i \gamma^\mu \psi_i
\rangle
= - {\delta \widetilde {\cal L}_{eff}\over \delta A_\mu} \quad ,\eqno(4.7)
$$
with $i= 1, 2$, where $\psi_1 = \psi $ and $\psi_2 = \chi$. Now
$$
q  j^\mu  = q \langle \overline \psi_i \gamma^\mu \psi_i \rangle
= \partial _\nu F^{\mu\nu} + \Complex \epsilon ^{\mu\nu\rho}F_{\nu\rho} \quad ,
\eqno(4.8)
$$
where $\Complex = {q^2\mu\over 4\pi \sqrt{M^2+\mu^2}}$ is the Chern--Simons
coefficient. We rewrite the last equation as
$$
qj^\mu = \left( \dal g^{\mu\nu} + 2 \Complex \partial _\rho \epsilon
^{\mu\nu\rho}\right)A_\nu\quad .\eqno(4.9)
$$

In momentum space, we have
$$
qj^\mu = -\left(p^2 g^{\mu\nu} + 2i \Complex p _\rho \epsilon
^{\mu\nu\rho}\right)A_\nu\quad .\eqno(4.10)
$$

Now, we write the $A_\mu$ field in terms of the current $\overline \psi_i
\gamma^\mu\psi_i$ inverting the above expression, then
$$
A_\mu = q \left[\left( {1\over 4\Complex ^2-p^2}\right) g_{\mu\rho} + {2i
\Complex p^\nu\over p^2 (4\Complex ^2 - p^2)} \epsilon _{\mu\nu\rho} +
p_\mu p_\rho \right] \left( \overline \psi _i \gamma^\rho \psi_i \right)\quad .
\eqno(4.11)
$$

Substituting (4.11)  in (4.6), we obtain
$$
\eqalignno{
\widetilde {\cal L}_{eff} &= \, i \overline \psi \gamma^\mu \partial_\mu
\psi +i\overline\chi\gamma_\mu\tilde\partial^\mu\chi + M(\overline \chi
\psi + \overline\psi \chi) + \mu
(\overline \psi \psi - \overline \chi\chi) \cr
& + q^2 (\overline \psi \psi - \overline \chi\chi) \dal^{-1}
(\overline \psi \psi - \overline \chi\chi) + q^2
(\overline \psi_i \gamma^\mu \psi_i) M_{\mu\nu}
(\overline \psi_j \gamma^\nu \psi_j)\quad ,&(4.12)\cr}
$$
where
$$
M_{\mu\nu} = {1\over 4\Complex ^2 - p^2} \left( g_{\mu\nu} - 2i\Complex {p^\rho
\over p^2} \epsilon _{\mu\nu\rho}\right)\quad .\eqno(4.13)
$$

If we are now interested in the low-momentum behaviour of the theory, we must
consider at small momentum $(p \to 0)$ each term of eq. (4.12). Thus, we can
write the low-momentum effective Lagrangian as
$$
\eqalignno{
\widetilde {\cal L}_{eff}& =  i (\overline \psi \not \!\!\partial \psi +
\overline \chi \tilde \not \!\!\partial \chi) + M(\overline \chi \psi +
\overline\psi \chi) + \mu (\overline \psi \psi - \overline \chi\chi) +\cr
& + q^2 (\overline \psi \psi - \overline \chi\chi) \dal^{-1}
(\overline \psi \psi - \overline \chi\chi) + {q^2 \over 4\Complex^2 - p^2}
(\overline \psi \gamma^\mu \psi + \overline \chi \gamma^\mu \chi)
(\overline \psi \gamma_\mu \psi + \overline \chi \gamma_\mu \chi)
\, .&(4.14)\cr}
$$

Notice that the above effective Lagrangian displays an interesting structure.
It contains two types of fermions besides non-local and quartic terms.

We believe that this new effective Lagrangian opens up a promising possibility
of getting a good description of anyonic superconductivity and
superfluity. It is also important to notice that the current (4.8)
calculated in terms of the Chern--Simons term can be used to get the Hall
conductivity of the vacuum, i.e.
$$
j^0 \sim \Complex \epsilon ^{0ij}F_{ij} = 2 \Complex F_{23} = 2 \Complex B\quad
,\eqno(4.15)
$$
whereas
$$
\eqalignno{
j^2 & \sim \, 2 \Complex \epsilon ^{203} F_{03} = i \sigma _{23}E^2\quad ,
&(4.16a)\cr
j^3 & =  \, 2 \Complex \epsilon ^{302} F_{02} \quad . &(4.16b)\cr}
$$

In the limit $\mu\to\infty$ we find $\sigma = {q^2\over 2\pi}$.

\vskip 1cm
\penalty-300
\centerline {\bf 5. Conclusions}
\vskip .7cm
\nobreak
In this article we have investigated the implementation of the KKB
scheme within the framework of the quantum electrodynamics in four dimensions.
We performed a dimensional reduction, that is, a compactification of an extra
spatial coordinate and obtained a reduced theory, the QED$_3$, which  exhibits
an interesting structure. We shown that the system described for this  model
undergoes a phase transition and, in addition, in the broken phase,
in which A field acquires mass, a Chern--Simons term survives. Moreover
We are also naturally led to fractional statistics since the Thirring
interaction can be bosonized in three dimensions\ref{19}. Actually, this fact
opens up a possibility
of arriving at similar conclusions to those drawn for $(1+1)$-dimensional
models, such as the chiral Gross--Neveu model where a bound state structure
exists, such that fermions are solitons formed as bound states of other fields
of the theory, leading to possible solution to fractional quantum Hall effect.
We have calculated the vector current for the model and have used it to get a
new effective Lagrangian in terms of the fermions fields of the theory. The
low-momentum behavior  of the Lagrangian is such that it describes some
special four-fermions interactions. This is a good indication of its
applicability in the theoretical studies on the basic phenomena of
anyonic superconductivity and Quantum Hall Effect as used in Ref. [18].
The vector current associated to the two types of fermions of the theory was
also used to get a Hall conductivity of the vacuum. The non-relativistic
quantum mechanics and the nature of the fermion-fermion interactions of
the KKB-QED$_3$ model are being presently investigated . Some results in this
direction have been found recently considering the issue of symmetry breaking
in a Gross--Neveu model with Thirring interaction.\ref{20}

\vskip 1cm
\centerline {\bf Acknowledgements}
\vskip .5cm
    One of the authors, F.M.C.F.,
would like to thank the Center for Theoretical Physics at M.I.T.,
while E.A. wishes to thank the International
Center for Theoretical Physics, Trieste, Italy, for their kind hospitality.
This work was also supported in part by CAPES  (Brazil).

\vskip 1cm
\penalty-300
\centerline {\bf References}
\vskip .7cm
\nobreak

\refer[[1]/S. Deser, R. Jackiw and S. Templeton, Ann. Phys. {\bf 140} (1982)
373;]

\refer[/R. Jackiw, Phys. Rev. {\bf D29} (1984) 2375]

\refer[[2]/C.R. Hagen, Ann. Phys. {\bf 157} (1984) 342; Phys. Rev. {\bf D36}
(1987) 3773]

\refer[[3]/E. Abdalla and F.M. de Carvalho Filho, Int. J. Mod. Phys. {\bf A7}
(1992) 619]

\refer[[4]/N. Redlich, Phys. Rev. Lett. {\bf 52} (1984) 18; Phys. Rev.
{\bf D29} (1984) 2366]

\refer[[5]/A.M. Polyakov, Mod. Phys. Lett. {\bf A3} (1988) 325;]

\refer[/I. Dzyaloshinnski, A.M. Polyakov and P. Wiegman, Phys. Lett.
{\bf 127A} (1988) 112]

\refer[[6]/F. Wilczek and A. Zee, Phys. Rev. Lett. {\bf 51} (1983) 2250;]

\refer[/Y. Wu and A. Zee, Phys. Lett. {\bf 147B} (1984) 325]

\refer[[7]/J.D. Lykken and J. Sonnenshein and N. Weiss, Int. J. Mod. Phys.
{\bf A6} (1991) 5155]

\refer[[8]/F. Wilczek, {\it Fractional Quantum Statistics and Anyon Super
Conductivity}, World Scientific, 1990]

\refer[[9]/X.G. Wen and A. Zee, Phys. Rev. {\bf B41} (1990) 240]

\refer[[10]/E. Abdalla, M.C.B. Abdalla and K. Rothe, {\it Non-perturbative
Methods in Two-\-dimen\-sional Quantum Field Theory}, World Scientific, 1991]

\refer [[11]/R.B. Laughlin, Phys. Rev. {\bf B23} (1981) 5632-5633]

\refer[[12]/B. Halperin, Phys. Rev. {\bf B25} (1982) 2185]

\refer[[13]/R.E. Prange and S.M. Girvin, ed., {\it The Quantum Hall Effect},
Springer, Berlin, 1987;]

\refer[/T. Chkraborty and P. Pietilainen {\it The Fractional Quantum Hall
Effect}, Springer-Verlag, 1988]

\refer[[14]/T. Kaluza, Sitzungsber. Preuss. Akad. Wiss. Phys. Math. K1 (1921)
966;]

\refer[/O. Klein, Z. Phys. {\bf 37} (1926) 895]

\refer[[15]/E. Cremmer and J. Scherk, Nucl. Phys. {\bf B118} (1977) 61]

\refer[[16]/A. Salam and J. Strathdee, Ann. Phys. {\bf 141} (1982) 316]

\refer[[17]/D.J. Toms, {\it An Introduction to Kaluza-Klein Theories},
H.C. Lee (ed.), World Scientific, Singapore, 1984]

\refer[[18]/E. Abdalla and M.C.B. Abdalla, CERN-TH 7344/94, cond-mat/9406119]

\refer[[19]/R. Banerjee, HD-THEP-95-17, hep-th/9504130]

\refer[[20]/T.S. Kim, W.H. Kye and J. K. Kim, preprint KAIST-TPP-K20,
hep-th/9509068]

\end